\documentclass[twocolumn]{apex}
\usepackage{graphicx}
\usepackage{amssymb}
\graphicspath{{images/}}

\title{Ferromagnetic SrRuO$_3$ thin-film deposition on a spin-triplet superconductor Sr$_2$RuO$_4$ with highly conducting interface}
\author{M. S. Anwar$^{1*}$, Y. J. Shin$^{2,3}$, S. R. Lee$^{2,3}$, S. J. Kang$^{2,3}$, Y. Sugimoto$^{1}$, S. Yonezawa$^{1}$, T. W. Noh$^{2,3}$, and Y. Maeno$^{1}$} 
\inst{$^{1}$Department of Physics, Graduate School of Science, Kyoto University, Kyoto 606-8502, Japan\\ $^{2}$Center for Correlated Electron Systems, Institute for Basic Science (IBS), Seoul 151-747, Republic of Korea\\ $^{3}$Department of Physics and Astronomy, Seoul National University, Seoul 151-747, Republic of Korea\\$^{*}$E-mail: anwar@scphys.kyoto-u.ac.jp;m.s.anwar476@gmail.com}

\date{\today}

\abst{Ferromagnetic SrRuO$_3$ thin films are deposited on the {\it ab}-surface of single crystals of the spin-triplet superconductor Sr$_2$RuO$_4$ as substrates using pulsed laser deposition. The films are under a severe in-plane compressive strain. Nevertheless, the films exhibit ferromagnetic order with the easy axis along the {\it c}-direction below the Curie temperature of 158 K. The electrical transport reveals that the SrRuO$_3$/Sr$_2$RuO$_4$ interface is highly conducting, in contrast with the interface between other normal-metals and the {\it ab}-surface of Sr$_2$RuO$_4$. Our results will stimulate the investigations on proximity effects between a ferromagnet and a spin-triplet superconductor.}

\begin{document}
\maketitle

Hybrid structures with ferromagnetic metals (FMs) and superconductors are fascinating systems exhibiting unconventional phenomena originating from competition and cooperation between the ferromagnetic order and superconductivity. In the past, two aspects of exotic behavior of the proximity effect between a FM and a spin-singlet superconductor (SSC) have been revealed. Firstly, the amplitude and phase of the spin-singlet order parameter spatially oscillates in the FM due to exchange interaction and even leads to a $\pi$-junction \cite{Ryazanov2001}. The penetration length of this effect is intrinsically limited to the order of a few nanometers. Secondly, a spin-triplet pair amplitude emerges over a length of the order of a micrometer into the FM if suitable magnetic inhomogeneity is present at the FM/SSC interface \cite{Bergeret2001,Keizer2006,Khaire2010,Robinson2010,Anwar2010,Anwar2012}. The resulting superconducting symmetry in the FM is spin-triplet {\it s}-wave and thus classified as odd-frequency pairing \cite{Tanaka2007}. To induce such a superconductivity into a ferromagnet, half metallic FM CrO$_2$ with multi-fold magnetic anisotropy \cite{Keizer2006,Anwar2010}, or alternatively ferromagnetic multilayer {\it X}/Co/{\it X}, where {\it X} can be PdNi, CuNi, Ni, or Ho, with non-collinear magnetization, is used in the SSC/FM/SSC Josephson junctions \cite{Khaire2010,Robinson2010,Anwar2012}.

Another approach to realize novel superconducting junctions with FM is to use a spin-triplet superconductor (TSC) with multiple degrees of freedom of the order parameter \cite{Gentile2013,Brydon2011,Linder2011}. Indeed, it has been theoretically predicted that not only charge supercurrent but also spin supercurrent emerges at FM/TSC interface and both can be controlled by the magnetization direction in the FM relative to the direction of the spins of the spin triplet Cooper pairs \cite{Brydon2011}. Thus, FM/TSC junctions, with both controllable charge and spin supercurrents, would initiate a new research area that can be termed as ``Superspintronics''.

Most likely, Sr$_2$RuO$_4$ (SRO214) exhibits the chiral {\it p}-wave spin-triplet superconducting order parameter analogous to the superfluid $^3$He-A phase \cite{Maeno2012,Maeno1994}. SRO214 has also been attracting much attention as one of the leading candidates for the topological superconductor. Recently, its topological nature has been investigated with superconducting junctions \cite{Anwar2013,Nakamura2011,Kashiwaya2011}. The main challenge in realizing a FM/TSC junction is the availability of thin films of spin triplet superconductors. By overcoming the difficulty in growing high quality SRO214 films, superconductivity has finally been induced in films \cite{Krockenberger2010}. Nevertheless, the transition temperature ($T_{\rm c}$) is about 1 K, which is noticeably lower than its bulk $T_{\rm c}$ of 1.5 K. Thus, as another approach to fabricate the FM/TSC junctions, we choose to use a high quality single crystal of SRO214, which is already available. SRO214 crystals can be cleaved along the {\it ab}-surface with atomically flat surface. However, there are surface reconstructions associated with RuO$_6$ octahedral rotation about the {\it c}-axis, which freezes the associated bulk soft phonon mode into a static lattice distortion. The reconstruction alters the surface electronic state \cite{Matzdorf2000} and in most cases leads to a poor electrical contact to the {\it ab}-surface. Therefore, it is an important issue to obtain highly conducting interface between a SRO214 crystal and a FM-layer in order to realize FM/TSC junctions. Here, we report the first growth of ferromagnetic epitaxial thin films of SRO113 on the {\it ab}-surface of single crystals of SRO214. Transport measurements reveal that the SRO113/SRO214 interface is highly conducting. 

SRO113 is an itinerant ferromagnet with the Curie temperature $T_{\rm Curie}$ of 160 K \cite{Koster2012}, and is the $n=\infty$ member of the same Ruddlesden-Poppor series Sr$_{n+1}$Ru$_n$O$_{3n+1}$ as SRO214 with $n=1$. SRO113 thin films have already been utilized in superconducting devices with the high temperature superconductor YBa$_2$Cu$_3$O$_{7-\delta}$ (YBCO). Interestingly, the supercurrent in the YBCO/SRO113/YBCO junction abruptly disappears when the thickness of the SRO113 layer exceeds 25 nm \cite{Char1992,Char1993}. In another report, superconducting gap in a YBCO/SRO113(18-nm) bilayer persists only at the ferromagnetic domain-wall regions \cite{Asulin2006}. Our findings of highly conducting interface between SRO113 and SRO214 open up a possibility to develop high-quality FM/TSC junctions to explore the unconventional proximity effects.

The {\it a}-axis mismatch at 300 K between the bulk SRO113 ($a_{\rm 113}=3.93$ \AA~in the pseudo-cubic notation) \cite{Chakoumkos1998} and the bulk SRO214 ($a_{\rm 214}=3.871$ \AA) \cite{Maeno1994} is $(a_{\rm 113}-a_{\rm 214})/a_{\rm 214}= 1.5$$\%$. This raises a possibility that a SRO113 (001) thin film can be grown on the {\it ab}-surface of SRO214 under compressive strain. Through out this paper, we use pseudo-cubic notation for the lattice parameters and crystalline axes of SRO113. We used SRO214 single crystals with a typical size of 3 $\times$ 3 $\times$ 0.5 mm$^3$  as substrates to grow SRO113 thin films using the pulsed laser deposition (PLD) technique with substrate-to-target distance of 54 mm. The growth temperature of 500 - 700$^\circ$C and the oxygen partial pressure of 100 mTorr were employed with the base pressure of $<$3 $\times$ 10$^{-7}$ Torr. It takes about 20 - 50 sec to grow one unit cell at a laser intensity of 1.5 Jcm$^{-2}$ and a repetition rate of 1 - 3 Hz. We also grow SRO113 thin film on an insulating SrTiO${_3}$ substrate as reference sample, whose residual resistivity ratio (RRR) is approximately 7, manifesting good performance of our growth facilities.  
%The thickness and quality of the film was monitored {\it in-situ} by the reflection high-energy electron diffraction (RHEED). X-ray diffraction (XRD) was used to examine the crystallography and morphology was studied with an atomic force microscope (AFM) of the films. 
The magnetization was measured down to 4 K with a SQUID magnetometer (Quantum Design, MPMS-XL). 

To measure the resistivity down to 4~K, we made a 1 $\times$ 0.7 mm$^2$ pad of SRO113 film and used a $^4$He cryostat (Quantum Design; PPMS). The cleaved {\it ab}-surface of SRO214 usually consists of regions of the typical size of 100 $\times$ 1000 $\mu$m$^2$ that look flat by optical microscope. Indeed, within such flat regions, we can easily find atomically flat areas of the size of 10 $\times$ 10 $\mu$m$^2$ under atomic force microscope (AFM). However, we also find several steps higher than the thickness of the film (50 nm) with a scanning electron microscope (SEM). If silver paste touches such a higher step, a direct contact between silver paste and SRO214 through the {\it ac}-surface would be created. To avoid such a scenario, we carefully put contacts only at most flat surfaces under optical microscope and absence of the direct contact was confirmed under SEM after measurements.

\begin{figure}[h]
		\begin{center}
\includegraphics[width=8cm]{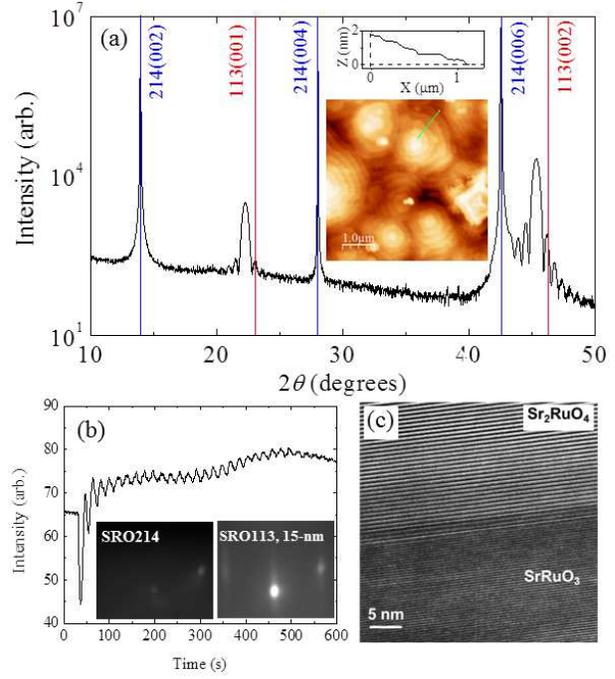}
\caption{(a) XRD spectrum on a logarithmic scale of a {\it c}-axis oriented 15-nm-thick SRO113 film deposited on a SRO214 substrate. Vertical lines indicate the positions of the corresponding peaks of the bulk materials. Insets show an AFM topographic image giving the average roughness of $\approx$ 0.5~nm and an atomic-step structure along the line given in the AFM-image. (b) RHEED oscillations during the film growth. Insets show the RHEED patterns before and after the film growth. (c) HR-TEM image of SrRuO$_3$$/$Sr$_2$RuO$_4$ hybrid. It is taken from [120] SrRuO$_3$$\parallel$[120] Sr$_2$RuO$_4$ zone axis.}
	\label{Figure1}
	\end{center}
\end{figure}

Figure \ref{Figure1}(a) represents the X-ray diffraction (XRD) spectrum of a 15-nm-thick SRO113 film grown on SRO214. It shows no impurity peaks but only the $(00l)$ peaks for SRO113 and SRO214, indicating that the ${\it c}$ axis of SRO113 film is in the same direction as that of SRO214 substrate. It is noted that the SRO113 $(00l)$ peaks are located at smaller angles than those of the bulk (vertical red solid lines), indicating that our film has an elongated lattice along the out-of-plane direction. The out-of-plane lattice constant is estimated as 4.00~\AA~in the pseudo-cubic notation, which is $1.8\%$ larger than the corresponding bulk value. AFM topographs of a film shown in the inset of Fig. \ref{Figure1}(a) reveal that the film has atomically-flat terraces with the step height equal to the lattice constant. Around the (001) and (002) peaks of SRO113, we find the thickness fringes (as the period of the oscillations well corresponds to the thickness of the films), which also suggest that the atomically flat top and bottom surfaces of SRO113 are formed.

In addition, the reflection high-energy electron diffraction (RHEED) patterns of the SRO113 film and the SRO214 substrate with the e-beam aligned along the [100] azimuthal direction are shown in the insets of Fig. \ref{Figure1}(b). Orientation and distance of the first-order peaks with respect to the central peaks are invariant before and after the SRO113 film deposition. This invariance indicates that the in-plane crystalline axes of the film and substrate are aligned and thus provide evidence for the epitaxial nature of the film. It is noted that the RHEED oscillations taken {\it in-situ} during the film growth as shown in Fig. \ref{Figure1}(b) demonstrates the changeover of the film-growth modes from layer-by-layer to step-flow at around 500 sec~\cite{Choi2010}. We also study SRO113/SRO214 hybrid using high resolution transmission electron microscopy (HR-TEM), which reveals that SRO113 films are grown epitaxially. 
%JEOL 2100F is used for the image acquisition and focused ion beam equipment is used for the sample preparation. 

Epitaxial growth of SRO113 on various perovskite substrates has been extensively studied: \cite {Choi2010,Shai2013} under compressive strain along the in-plane direction, SRO113 films exhibit the {\it c}-axis elongation. We found that our SRO113 film has essentially the same peak positions as the one deposited on an NdGaO$_3$ substrate \cite{Shai2013}, which has an in-plane lattice constant (3.86~\AA~in the pseudo-cubic notation) very similar to that of SRO214.

\begin{figure}[h]
		\begin{center}
\includegraphics[width=8cm]{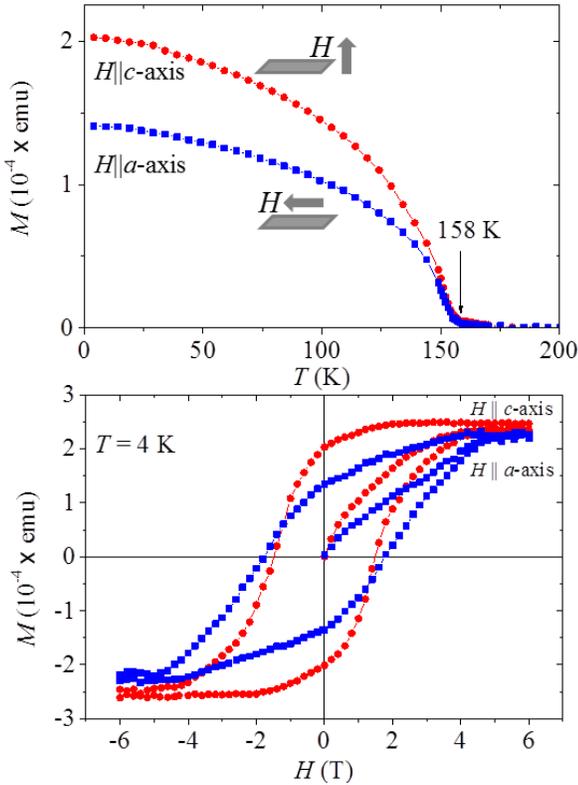}
\caption{(a) Temperature dependent remanent magnetization of a 50-nm-thick SRO113 film along both the {\it a}-axis (blue squares) and the {\it c}-axis (red circles). The sample was cooled down in field of 1~T and its magnetization was measured upon warming after removing field at 4 K. (b) Magnetization loops of the film along the {\it a}- and {\it c}-axes. In both (a) and (b), the substrate contributions were measured before deposition of the film and were subtracted from data measured after deposition.}
	\label{Figure2}
	\end{center}
\end{figure}

Anticipating that the strong compressive strain could modify the magnetic behavior of the film, we measured its magnetization. Figure \ref{Figure2} (a) shows the temperature dependence of the magnetization of a 50-nm-thick SRO113 film with the field of 10~mT both along the {\it c}-axis (out-of-plane) and the {\it a}-axis (in-plane) directions on field cooling. Surprisingly, we observe that $T_{Curie}$ is 158~K, which is almost equal to the values for the bulk \cite{Gan1998,Luo2007}. In contrast, it is reported that high-quality SRO113 thin films deposited on SrTiO$_3$ substrates show the reduction of $T_{\rm Curie}$ to 150 K, although the {\it a}-axis mismatch between SRO113 and SrTiO$_3$ is only about $-0.45\%$ \cite{Koster2012}. After careful subtraction of the background signals, we obtain the remanent magnetization of the film at 4~K to be 2.8 $\mu_{\rm B}$/Ru along the {\it c}-axis and 2 $\mu_{\rm B}$/Ru along the {\it a}-axis, which are substantially larger than the expected values \cite{Kambayasi1976}. The magnetization loops for both the {\it c}- and {\it a}-axis direction are presented in Fig. \ref{Figure2} (b). These data show that our films exhibit a strong magnetic anisotropy with larger remanent magnetization along the {\it c}-axis \cite{Grutter2012}. We obtained essentially the same $T_{\rm Curie }$, magnetization values, and anisotropy for a 15-nm-thick SRO113 film.

\begin{figure}[h]
		\begin{center}
\includegraphics[width=8cm]{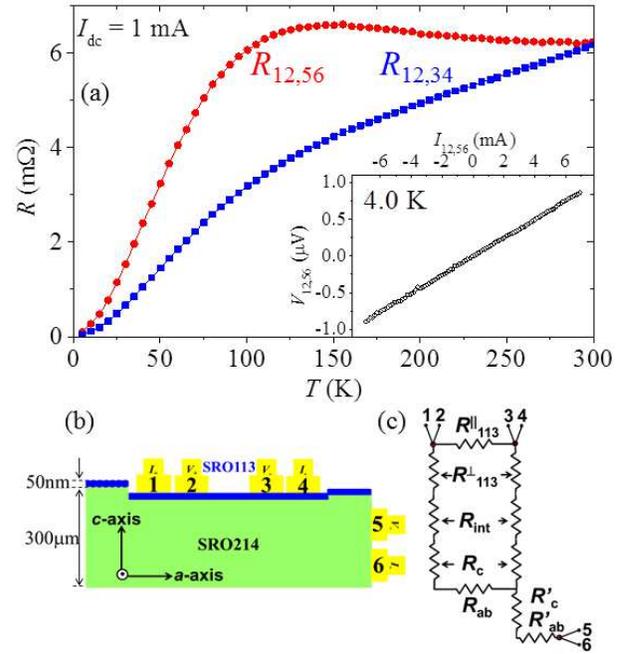}
	\caption{Resistance vs temperature of a SRO113(50 nm)/SRO214 hybrid system. (a) $R_{12,34}(T)$ (on the {\it ab}-surface; blue squares) and $R_{\rm 12,56}$($T$) (top to side; red circles). Inset shows a linear {\it I}-{\it V} curve measured in 12,56 configuration. (b) Schematic illustration of the electrodes. Top electrodes 1 to 4 are placed in a flat area away from the steps larger than the film thickness. (c) Simplified model of the resistance circuit.}
	\label{Figure3}
	\end{center}
\end{figure}

Figure \ref{Figure3} (a) presents the temperature dependent resistance $R$($T$) between 300 K and 4 K for a 50-nm-thick SRO113 film. We measured $R$($T$) using a dc four-probe technique with two different channel configurations: first with four contacts on the SRO113 film $R$$_{\rm 12,34}$($T$); second with two electrodes on SRO113 film and the other two electrodes on the side of SRO214 substrate $R$$_{\rm 12,56}$($T$). A schematic of the sample with the electrode arrangements is shown in Fig. \ref{Figure3} (b). Note that the electrodes were placed in a flat area, avoiding steps larger than film thickness. The characteristic slope change around 120 K for both data suggests that the {\it c}-axis resistivity $\rho_{\rm c}$ of SRO214 \cite{Hussey1998} contributes significantly to transport for both configurations. It is obvious that $\rho_{\rm c}$ contributes more for $R$$_{\rm 12,56}$($T$) as expected from the electrode configuration. Thus, $R$($T$) mainly reflects the behavior of SRO214 in the whole temperature range down to 4 K. This fact suggests that the SRO113 film has a good electrical contact with the SRO214 substrate. A linear current-voltage ({\it I}-{\it V}) curve measured at 4~K with 12,56 configuration of the electrodes also reveals that we obtain an Ohmic contact between SRO113 and SRO214. Note that the residual resistance ratio (RRR) for 12,34 configuration is 60  which is many folds larger than that of SRO113 thin films, and much smaller than that of SRO214. This RRR value suggests small but non-negligible resistance contribution of the interface in particular at low temperatures.   

%Figure \ref{Figure3} (b) shows $R$($T$) around the $T_c$ of SRO214. The data show the transition to zero resistance around 1.2 K at $I$$_{\rm dc}$= 1 mA, indicating again the contribution from the bulk SRO214 is dominant even at the low temperatures. 

To further discuss the resistance of the SRO113/SRO214, we use a simplified model circuit shown in Fig. \ref{Figure3} (c). In this circuit, $R^{\parallel}_{113}$ and $R^{\bot}_{113}$ are the resistances of the SRO113 film parallel and perpendicular to the {\it ab}-surface, $R_{\rm int}$ is the interface resistance, and $R_c$ and $R_{ab}$ are the resistances of SRO214 along the {\it c}-axis and along the {\it ab}-plane. We should note that this simplified model estimates the upper limits for the measured resistances. Thus, the measured values should be lower than the estimated values. In order to estimate each resistance, we refer to existing resistivity data of SRO113 and SRO214. At 300~K, SRO113 thin films exhibit nearly isotropic resistivity $\rho_{\rm 113}$ $\approx$ 250 $\mu$$\Omega$cm depending on the substrate \cite{Koster2012}. For any kind of substrate, $\rho_{\rm 113}$ decreases linearly with decreasing temperature down to $T_{\rm Curie}$ and exhibits a sharp kink at $T_{\rm Curie}$ due to the reduction of scatterings by spin fluctuations. The resistivity of SRO214 along the {\it c}-axis is $\rho_{c}$ $\approx$ 15 m$\Omega$cm at 300~K, which is two orders of magnitude larger than the resistivity along the {\it ab}-surface $\rho_{ab}$ $\approx$ 120 $\mu$$\Omega$cm \cite{Hussey1998}. Whereas $\rho_{ab}$ decreases monotonically with decreasing temperature, $\rho_{c}$ exhibits a broad maximum around 120 K \cite{Hussey1998}. Using these resistivity values and the geometrical factors of the present hybrid system, we performed an order estimate of each contribution of the present system at 300~K: $R^{\parallel}_{113}$ $\approx$ 70 $\Omega$, $R^{\bot}_{113}$ $\approx$ 0.2 $\mu\Omega$, $R_c$ $\approx$ 10 m$\Omega$, and $R_{ab}$ $\approx$ 3 m$\Omega$. We also estimate each contribution at 4~K using the values of $\rho_{\rm 113} \approx$ 10 $\mu\Omega$cm, $\rho_{c}$ $\approx$ 1 m$\Omega$cm, and $\rho_{ab}$ $\approx$ 1 $\mu$$\Omega$cm: we obtain $R^{\parallel}_{113}$ $\approx$ 3 $\Omega$, $R^{\bot}_{113}$ $\approx$ 10 n$\Omega$, $R_c$ $\approx$ 1 m$\Omega$, and $R_{ab}$ $\approx$ 20 $\mu\Omega$.

In the present model, $R_{\rm 12,34}$ should satisfy the relation $1/R_{12,34}=1/R^{\parallel}_{113}+1/(2R^{\bot}_{113}+2R_{\rm int}+2R_c+R_{ab})$. Since $R^{\parallel}_{113}$ at 300 K is estimated to be four orders of magnitude larger than the observed value of $R_{\rm 12,34}=6~$m$\Omega$, the first term on the right hand side should be negligible. In the second term, because $R^{\bot}_{113}$ is estimated to be very small compared with $R_c$ or $R_{ab}$, the $R^{\bot}_{113}$ term has little contribution. Thus, the contributions of the SRO113 film to $R_{\rm 12,34}$ should be negligible. This is indeed consistent with the fact that resistance anomaly at $T_{\rm Curie}$ is absent in our results. Now, the relation can be reduced as $R_{\rm 12,34} \approx 2R_{\rm int}+2R_c+R_{ab}$. We can then notice that the estimated value of $2R_c+R_{ab}$ $\approx$ 20~m$\Omega$ at 300~K and 2~m$\Omega$ at 4~K and the observed $R_{\rm 12,34}$ value ($\approx$ 6~m$\Omega$) are on the same order. The observed temperature dependence of $R_{\rm 12,34}$ is also understood as a certain combination of $R_c$ and $R_{ab}$. Therefore, $R_{\rm int}$ should have only a very small contribution. For $R$($T$)$_{\rm 12,56}$, because contributions of $R^{\parallel}_{113}$ and $R^{\bot}_{113}$ are again ignorable, we obtain $R_{12,56} \cong R_{\rm int}+R_c+R_{ab}+R^{'}_c+R^{'}_{ab}$ $(R^{'}_c$ and $R^{'}_{ab}$ are additional bulk resistances), and the same conclusion is deduced. 

Our conclusion of the highly conducting SRO113/SRO214 interface apparently contradicts with previous experimental works indicating that the SRO214 {\it ab}-surface is not a good metal because of the surface reconstruction accompanied by the RuO$_6$ octahedral rotation \cite{Matzdorf2000}. Indeed, we usually cannot achieve good electrical contact between normal metals and the SRO214 {\it ab}-surface. In addition, such a surface reconstruction is thought to destroy the superconductivity in the {\it ab}-surface region of SRO214. Thus, the highly conducting interface in the present hybrid system is surprising. This observation indicates that the surface reconstructions might be suppressed under expansive surface strain on SRO214 caused by the epitaxial growth of SRO113 films. The detailed interface investigations are also interesting topics in the future.

To summarize, we grow ferromagnetic epitaxial SRO113 thin films by PLD using cleaved single crystals of superconducting SRO214 substrates. The films are under severe compressive strain but with very small reduction in $T_{\rm Curie}$ compared to the bulk. Resistivity measurements reveal that the interface between SRO113 and SRO214 is highly conducting. The epitaxial growth might relax SRO214 surface reconstructions and make the interface rather conducting. The SRO113/SRO214 hybrid system opens a possibility to study FM/TSC junctions in future.

\acknowledgement We are thankful to D. Manske, K. Char, S. B. Chung, M.R. Cho, and K. Lahabi for valuable discussions. We acknowledge technical support by M. P. Jimenez S. This work was supported by the "Topological Quantum Phenomena" (No. 22103002) KAKENHI on Innovative Areas from the Ministry of Education, Culture, Sports, Science and Technology (MEXT) of Japan, and also supported by the Institute for Basic Science (IBS) in Korea.


\begin{thebibliography}{99}
\bibitem{Ryazanov2001} V. V. Ryazanov, V. A. Oboznov, A. Yu. Rusanov, A. V. Veretennikov, A. A. Golubov, and J. Aarts, Phys. Rev. Lett. {\bf 86}, 2427 (2001).
\bibitem{Bergeret2001} F. S. Bergeret, A. F. Volkov, and K. B. Efetov, Phys. Rev. Lett. {\bf 86}, 4096 (2001).
\bibitem{Keizer2006} R. S. Keizer, S. T. B. Goennenwein, T. M. Klapwijk, G. Miao, G. Xiao, and A. Gupta, Nature {\bf 439}, 825 (2006).
\bibitem{Khaire2010} T. S. Khaire, M. A. Khasawneh, W. P. Pratt, Jr., and N. O. Birge, Phys. Rev. Lett. {\bf 104}, 137002 (2010).
\bibitem{Robinson2010} J. W. A. Robinson, J. D. S. Witt, and M. G. Blamire, Science {\bf 329}, 59 (2010).
\bibitem{Anwar2010} M. S. Anwar, F. Czeschka, M. Hesselberth, M. Porcu, and J. Aarts, Phys. Rev. B {\bf 82}, 100501(R) (2010).
\bibitem{Anwar2012} M. S. Anwar, M. Veldhorst, A. Brinkman, and J. Aarts, Appl. Phys. Lett. {\bf 100}, 052602 (2012).
\bibitem{Tanaka2007} Y. Tanaka, A. A. Golubov, S. Kashiwaya, and M. Ueda, Phys. Rev. Lett. {\bf 99}, 037005 (2007).
\bibitem{Brydon2009} P. M. R. Brydon and D. Manske, Phys. Rev. Lett. {\bf 103}, 147001 (2009).
\bibitem{Gentile2013} P. Gentile, M. Cuoco, A. Romano, C. Noce, D. Manske, and P. M. R. Brydon, Phys. Rev. Lett. {\bf 111}, 097003 (2013).
\bibitem{Brydon2011} P. M. R. Brydon, W. Chen, Y. Asano, and D. Manske, Phys. Rev. B {\bf 88}, 054509 (2013).
\bibitem{Linder2011} G. Annunziata, M. Cuoco, C. Noce, A. Sudbø, and J. Linder, Phys. Rev. B {\bf 83}, 060508(R) (2011).
\bibitem{Maeno2012} Y. Maeno, S. Kittaka, T. Nomura, S. Yonezawa, and K. Ishida, J. Phys. Soc. Jpn. {\bf 81}, 011009 (2012).
\bibitem{Maeno1994} Y. Maeno, H. Hashimoto, K. Yoshida, S. Nishizaki, T. Fujita, J. G. Bednorz, and F. Lichtenberg, Nature {\bf 372}, 532 (1994).
\bibitem{Anwar2013} M. S. Anwar, T. Nakamura,	S. Yonezawa,	M. Yakabe,	R. Ishiguro,	H. Takayanagi, and Y. Maeno, Sci. Rep. {\bf 3}, 2480 (2013).
\bibitem{Nakamura2011} T. Nakamura, R. Nakagawa, T. Yamagishi, T. Terashima, S. Yonezawa, M. Sigrist, and Y. Maeno, Phys. Rev. B {\bf 84}, 060512 (2011).
\bibitem{Kashiwaya2011} S. Kashiwaya, H. Kashiwaya, H. Kambara, T. Furuta, H. Yaguchi, Y. Tanaka, and Y. Maeno, Phys. Rev. Lett. {\bf 107}, 077003 (2011).
\bibitem{Krockenberger2010} Y. Krockenberger, M. Uchida, K. S. Takahashi, M. Nakamura, M. Kawasaki and Y. Tokura, Appl. Phys. Lett. {\bf 97}, 082502 (2010).
\bibitem{Matzdorf2000} R. Matzdorf, Z. Fang, Ismail, J. Zhang, T. Kimura, Y. Tokura, K. Terakura, and E. W. Plummer, Science {\bf 289}, 746 (2000).
\bibitem{Koster2012} G. Koster, L. Klein, W. Siemons, G. Rijnders, J. S. Dodge, C. B. Eom, D. H. A. Blank, and M. R. Beasley, Rev. Mod. Phys. {\bf 84}, 253 (2012).
\bibitem{Char1993} L. Antognazza, K. Char, T. H. Geballe, L. L. H. King and A. W. Sleight, Appl. Phys. Lett. {\bf 63}, 1005 (1993).
\bibitem{Char1992} K. Char, M. S. Colclough, T. H. Geballe, and K. E. Myers, Appl. Phys. Lett. {\bf 62}, 196 (1993).
\bibitem{Asulin2006} I. Asulin, O. Yuli, G. Koren, and O. Millo, Phys. Rev. B {\bf 74}, 092501 (2006).
\bibitem{Chakoumkos1998} B. C. Chakoumakos, S. E. Nagler, S. T. Misture, H. M. Christen, Physica B {\bf 241-243}, 358 (1998); the actual structure is orthorhombic with $a=5.537 $\AA, $b=7.852$ \AA, and $c=5.573$ \AA~at 300 K.  
\bibitem{Choi2010} K. J. Choi, S. H. Baek, H. W. Jang, L. J. Belenky, M. Lyubchenko, and C. B. Eom, Adv. Mater. {\bf 22}, 759 (2010). 
\bibitem{Shai2013} D. E. Shai, C. Adamo, D. W. Shen, C. M. Brooks, J. W. Harter, E. J. Monkman, B. Burganov, D. G. Schlom, and K. M. Shen, Phys. Rev. Lett. {\bf 110}, 087004 (2013).
\bibitem{Gan1998} Q. Gan, R. A. Rao, C. B. Eom, J. L. Garrett and M. Lee, Appl. Phys. Lett. {\bf 72}, 978 (1998).
\bibitem{Luo2007} H. M. Luo, M. Jain, S. A. Baily, T. M. McCleskey, A. K. Burrell, E. Bauer, R. F. DePaula, P. C. Dowden, L. Civale, and Q. X. Jia, J. Phys. Chem. B {\bf 111}, 7497 (2007).
\bibitem{Kambayasi1976} A. Kanbayasi, J. Phys. Soc. Jpn. {\bf 41}, 1876 (1976).
\bibitem{Grutter2012} A. J. Grutter, F. J. Wong, E. Arenholz, A. Vailionis, and Y. Suzuki, Phys. Rev. B {\bf 85}, 134429 (2012).
\bibitem{Hussey1998} N. E. Hussey, A. P. Mackenzie, J. R. Cooper, Y. Maeno, S. Nishizaki and T. Fujita, Phys. Rev. B {\bf 57}, 5505 (1998).
\end{thebibliography}
\end{document}